# Human sensory-musculoskeletal modeling and control of whole-body movements

Chenhui Zuo[1†], Guohao Lin[1†], Chen Zhang[1†], Shanning Zhuang[1], Yanan Sui[1*]

Tsinghua University.

*Corresponding author. E-mail: ysui@tsinghua.edu.cn.
†These authors contributed equally to this work.

## Abstract

Coordinated human movement depends on the integration of multisensory inputs, sensorimotor transformation, and motor execution, as well as sensory feedback resulting from body-environment interaction. Building dynamic models of the sensory-musculoskeletal system is essential for understanding movement control and investigating human behaviours. Here, we report a human sensory-musculoskeletal model, termed SMS-Human, that integrates precise anatomical representations of bones, joints, and muscle-tendon units with multimodal sensory inputs involving visual, vestibular, proprioceptive, and tactile components. A stage-wise hierarchical deep reinforcement learning framework was developed to address the inherent challenges of high-dimensional control in musculoskeletal systems with integrated multisensory information. Using this framework, we demonstrated the simulation of three representative movement tasks, including bipedal locomotion, vision-guided object manipulation, and human-machine interaction during bicycling. Our results showed a close resemblance between natural and simulated human motor behaviours. The simulation also revealed musculoskeletal dynamics that could not be directly measured. This work sheds deeper insights into the sensorimotor dynamics of human movements, facilitates quantitative understanding of human behaviours in interactive contexts, and informs the design of systems with embodied intelligence.

## Introduction

Humans interact with the external world through coordinated body movements enabled by the interplay of the sensory, musculoskeletal, and central nervous systems. These systems orchestrate multisensory integration, sensorimotor transformation, and motor execution in a closed-loop manner to ensure precise and adaptive movement control. Developing a comprehensive and fully controllable whole-body model of the human sensory-musculoskeletal system is important for facilitating computational analysis of coordinated motor control and to design humanoid systems with embodied intelligence.

Modeling the human sensory-musculoskeletal system requires a detailed representation of anatomical structures, biomechanical properties, and sensorimotor control processes. The adult human musculoskeletal system comprises approximately 206 bones, over 200 joints, and more than 600 skeletal muscles, collectively allowing a broad spectrum of motor actions. The precise counts of joints and muscles vary with the classification methods, as well as individual and gender variations[1]. The skeleton forms a rigid framework with joints facilitating articulation between bones. Skeletal muscles are attached to bones via tendons and generate forces by muscle contraction to execute motor actions. Previous human musculoskeletal models focused mainly on the control and biomechanics of specific parts of the body, such as the upper limbs for manipulation[2,3] and the lower extremities for walking[4–6]. While these models provided insights into joint mechanics and muscle forces, they often simplified muscle structures or omitted sensory components, limiting their utility for simulating whole-body coordinated movements. Some models allowed whole-body simulation but with simplified muscle construction to ease the control process [7,8], and thus could not fully simulate physiologically aligned human movements and muscle activations. Skeletal models provided insights into biomechanics [9,10], but failed to capture the rich spatiotemporal dynamics of muscle activations critical for generating naturalistic motor behaviours. Existing musculoskeletal modeling efforts also struggled with high-dimensional control [11,12], often failing to generate robust closed-loop behaviours that reproduce desired motions.

Beyond biomechanics, sensory feedback is essential for adaptive motor control. The proprioceptive system provides information on body segment positions and muscle states, while mechanoreceptors in the skin and tendons detect mechanical stimuli. The vestibular system senses head orientation and balance, and the visual system enables spatial awareness and object tracking. Integrating these sensory modalities is fundamental for maintaining dynamic stability and coordinating complex motor tasks.

The inherent complexity of the sensory-musculoskeletal system poses significant computational challenges. Muscle dynamics exhibit nonlinear relationships between force, length, and velocity, while multi-joint coordination further complicates control strategies. Moreover, the ultra-high dimensionality of muscle actuation space introduces redundancies, resulting in multiple muscle activation patterns for the same movement goal. The dynamic coordination of motor outputs in response to environmental changes also adds to the complexity of motor control during movement tasks. These challenges are not properly addressed in most existing models that did not incorporate sufficient multisensory inputs [13,14].

To overcome these challenges, we developed a sensory-musculoskeletal human model, termed SMS-Human, that integrates the whole-body musculoskeletal system with multiple sensory components. This model achieves comprehensive anatomical representation that comprises 175 rigid body segments with 206 precise bone meshes, 278 joints, and 1,266 muscle-tendon units, with accurate spatial arrangement and validated functional parameters. We incorporated multimodal sensory (binocular visual, vestibular, proprioceptive, and tactile) inputs (Fig. 1a-f) in our simulation with the open-source MuJoCo physics engine [15].

We developed a hierarchical deep reinforcement learning (DRL) approach with efficient representations of high-dimensional actuators and a stage-wise learning process for training a neural network controller that could simulate human motor behaviours using the SMS-Human model. We demonstrated successful control of coordinated human-like movements with high physiological and behavioural fidelity, including bipedal locomotion, vision-guided object manipulation, and bicycling. The unprecedented anatomical detail of our model provides a new foundation for computational analysis of human sensory-musculoskeletal system in a closed-loop, physics-based way, and for studying spatiotemporal dynamics for whole-body movement control.

## Results

### Overall scheme of sensory-musculoskeletal modeling and control

The SMS-Human model integrates multimodal sensory inputs to enable sensorimotor transformation for motor behaviours in the simulation environment (Fig. 1). Binocular sensing was implemented via the cameras positioned in the eyes, providing the egocentric view of the model during the task performance (Fig. 1a, b). The vestibular system was provided by sensors located in the head, measuring the linear and rotational head motion and orientation important for maintaining balance and coordinating eye-head movements (Fig. 1c). Proprioceptive components encoded the angular position (P), velocity (V), and acceleration (A) for all joints, as well as the length (L), velocity (V), force (F), and activation (Act) for all muscle-tendon units (Fig. 1d). This comprehensive proprioceptive information on the joints and muscle-tendon units allows precise control of posture and movement. We placed touch sensors for measuring contact forces over the body, as shown by the forces detected at the hands (Fig. 1e) and feet (Fig. 1f). This tactile feedback is important for the model to execute stable bipedal locomotion and fine motor tasks such as object manipulation.

Each modality of sensory input was represented by corresponding matrices shown in Fig. 1j. A neural network controller was trained to generate motor output in the action space based on multisensory input, as represented by the motor actuation matrix (Fig. 1k). The learned motor actions served as neural excitation to actuate skeletal muscles for motor behaviours in response to the observed state (Fig. 1h,i). Both the state and action spaces were high-dimensional in nature (Fig. 1j,k). This scheme of integrating biomechanical model with multimodal sensing represents the basis for the embodied simulation.

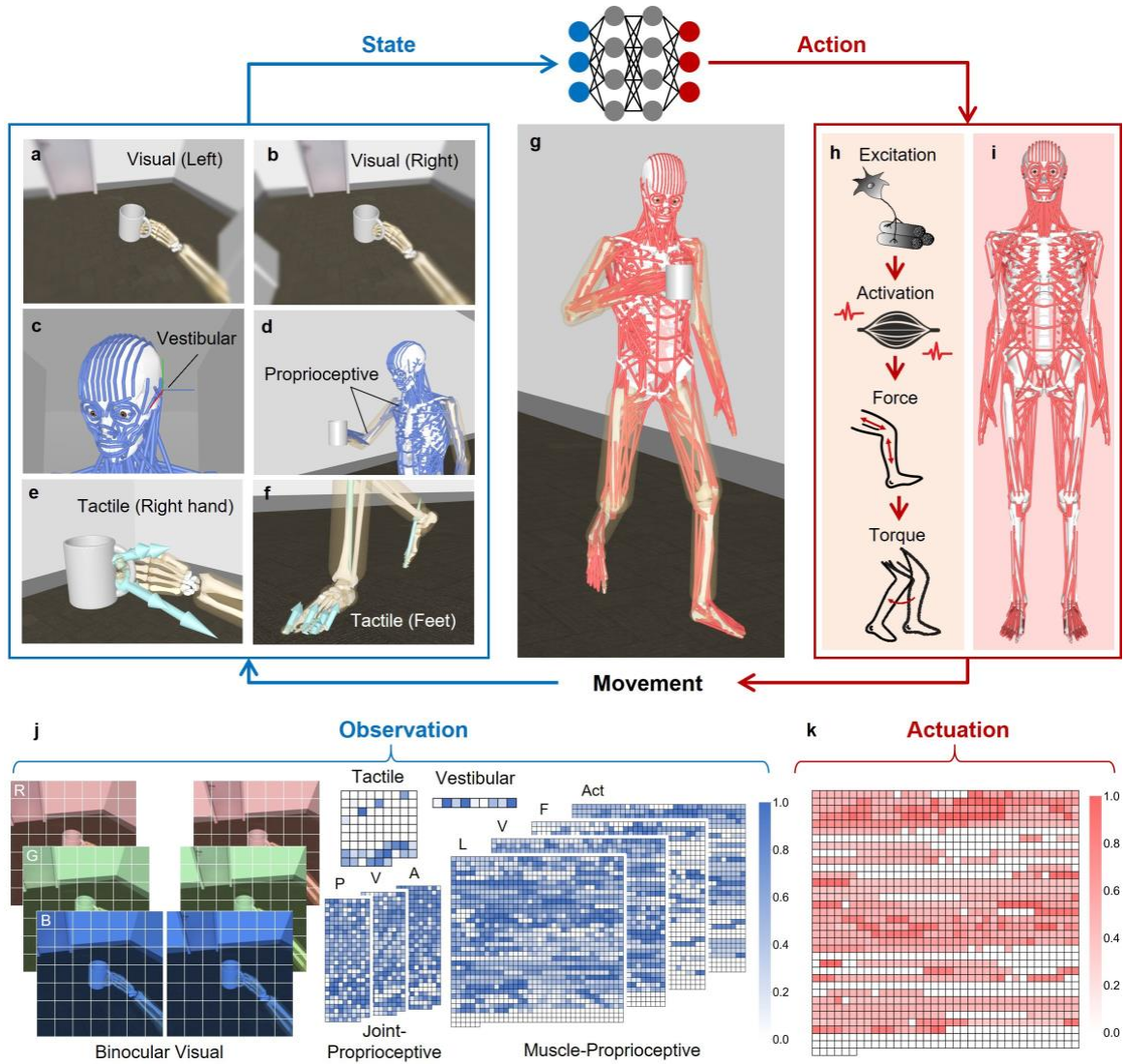

**Fig. 1 Human sensory-musculoskeletal model with sensorimotor transformation and feedback control**. **a,b**, Binocular inputs provided by cameras in the left eye (**a**) and right eye (**b**), with progressive peripheral blurring to simulate human foveal vision. **c**, Vestibular information provided by sensors located in the head. **d**, Proprioceptive inputs provided by sensors in joints and muscle-tendon units. **e,f**, Tactile inputs in hands (**e**) and feet (**f**), with light blue arrows indicating contact forces. **g**, Full-body visualization of SMS-Human during walking, while looking at the mug in the hand. **h**, Neural control pathway for movement generation, including neural excitation, muscle activation, force and torque generation. **i**, Musculoskeletal representation of SMS-Human model, featuring 1,266 muscle-tendon units (red lines) to actuate the whole-body skeleton. **j**, High-dimensional inputs to the DRL network at a single time point, including binocular images at RGB channels (224x224x3x2), conjugated vestibular inputs (9x1), tactile inputs (78x1), joint proprioceptive inputs (278x3), and muscular proprioceptive inputs (1266x4). Separate matrices for position (P), velocity (V), acceleration (A), length (L), force (F), and activation (Act) are shown for proprioceptive components. Color intensity represents the normalized value for all modalities except vision (represented in RGB space). **k**, Motor activation in the action space (1266x1). Color intensity represents activation level. In **a–g**, translucent yellow capsules surrounding skeletal structures represent the body-environment interface modeling.

## Construction of embodied human sensory-musculoskeletal model

To faithfully capture the biomechanical complexity of the human body, we decomposed anatomical structures of the human musculoskeletal system into functional units at an unprecedented scale of granularity. The 206 bones were grouped into 175 rigid body segments capable of independent movement and connected by 278 joints. Human skeletal muscles, which attach to bones and exert forces via tendons, were implemented as 1,266 muscle-tendon units. Muscles with broad attachment positions were modeled using multiple muscle-tendon units to enable more precise dynamic control. This modeling approach resulted in a greater number of muscle-tendon units than the anatomical skeletal muscle counts.

We constructed the skeleton for a typical adult male through selective integration of existing databases and models [2,3,5,16,17], with further refinement based on published anatomical data. The bone meshes in the model were simplified to optimize simulation efficiency, while preserving precise anatomical landmarks and bony features essential for accurate muscle attachments. The body segment-specific dynamic parameters included mass and inertial properties, which were calculated by estimating segment volume and tissue density. This fine-grained representation of the mass and inertia distribution allowed for a biomechanically more realistic simulation of body movements.

The body segments were articulated through joints, whose position, axis orientation, range of motion (ROM), and kinematic coupling were explicitly parameterized within the model. We determined the joint configurations by detailed examination of data from original human anatomical studies [18–23] and previous simulation models [2,17,24–26]. Special attention was paid to previously unmodeled joints and articulation constraints to achieve physiologically accurate ROM and naturalistic motion patterns during whole-body movements.

The muscular system was developed by rigorously following Gray's Anatomy [1], resulting in unprecedented completeness and precision in musculature representation. The path of each muscle-tendon unit was individually verified and refined to ensure anatomical accuracy in the routes and attachment points. The architectural parameters used in the simulation, including optimal fiber length, physiological cross-sectional area, pennation angle and tendon slack length were implemented based on existing anatomical studies and validated models [2–4,8,24,26–33]. For muscle parameters not included in existing literature, we developed a systematic approach for parameter estimation based on primary anatomical data[34–42]. The force-generating capability of a muscle-tendon unit was modeled with normalized fiber force-length-velocity relationship[43] and neuromuscular dynamics [44], ensuring physiologically plausible muscle behaviours. We excluded certain anatomically defined skeletal muscles that are not directly involved in joint articulation, including the diaphragm, pelvic floor muscles, and muscles involved in deglutition.

Figure 2 highlights representative muscle-tendon units across various body parts in our model, including those for controlling the eyeball, jaw (Fig. 2a), torso (Fig. 2b and 2c), neck (Fig. 2d), hand (Fig. 2f) and foot (Fig. 2g). These units were either absent or modeled with insufficient anatomical details in previously reported models. To ensure physiologically accurate muscle-tendon contractile trajectories, we incorporated wrapping surfaces around body segments. As shown in the example in Fig. 2e, these surfaces accounted for the physical constraints imposed by bones, deep muscles, and soft tissues.

Together, our SMS-Human model represents a musculoskeletal system with much higher anatomical accuracy than that of existing 3D anatomical databases[16,45]. The high dimensionality of muscle-tendon units poses both advantages and challenges as a platform for realistic simulation of human movements.

## Deep reinforcement learning for sensorimotor control

The high dimensionality of muscle actuation present a fundamental challenge in motor control[46,47]. In humans, motor control commands are generated via a series of sensorimotor processes and executed through coordinated muscle synergistic activation, enabling adaptability, efficiency, and robustness during movement[48,49]. Inspired by human motor control, we developed a stage-wise hierarchical deep reinforcement learning method to control the SMS-Human model to perform whole-body movement tasks (Fig. 3). It took as input the multimodal sensory information from the state space, generating control outputs for hierarchically represented muscles in the action space, with muscle grouping based on dynamic similarities and anatomical constraints. Within each muscle group, we implemented a hierarchical control strategy, where muscles received shared group control signals for group-level coordination and individual state-dependent refinements based on sensory feedback and task demands.

During training, visual, vestibular, proprioceptive, and tactile inputs were combined to form an integrated observation matrix of multisensory information. This matrix was then delivered to a Soft

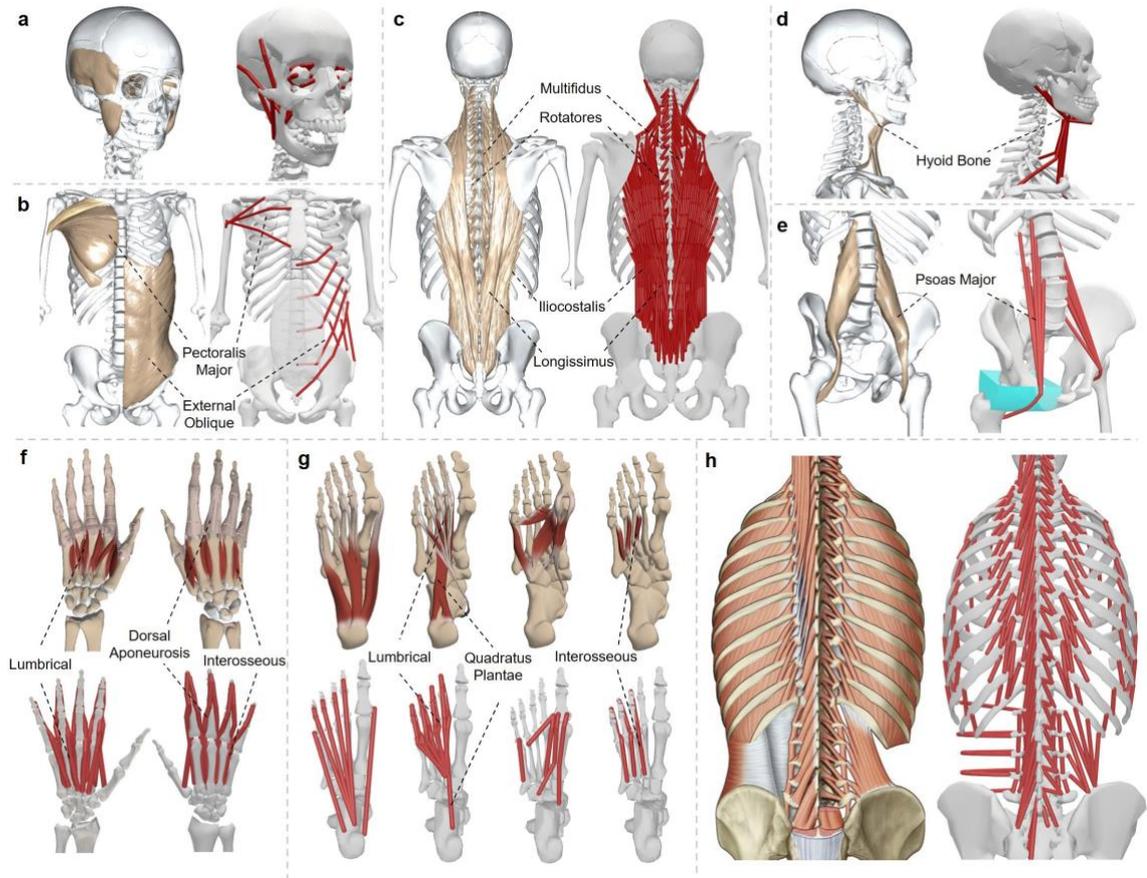

**Fig. 2 Anatomically precise reconstruction of muscle-tendon units in the SMS-Human model.** Side-by-side comparisons of the anatomical arrangements of representative muscles and corresponding muscle-tendon units (thick red lines) in the model. In **a-e**, BodyParts3D anatomical database [16] (left) was used for modeling muscle-tendon units (right). In **f-g**, Primal Pictures 3D atlas database [45] (upper) was used for modeling muscle-tendon units (lower). **a**, Head, temporomandibular joint, masticatory and extraocular muscles, and corresponding muscle-tendon units. **b**, Thorax and abdominal region, pectoralis major and external oblique muscles and corresponding muscle-tendon units. **c**, Back region, iliocostalis and longissimus muscles were modeled with multiple muscle-tendon units. Multifidus and rotatores muscles between spinal segments were modeled individually. **d**, Neck, the hyoid bone and the attached suprahyoid/infrahyoid muscle groups and corresponding model units. **e**, Wrapping surfaces (shown in light blue) were added around body segments for constraining the muscle contractile pathways, as illustrated here for the psoas major muscle. **f**, Hand, palmar (left) and dorsal (right) views of the right hand. Lumbrical and interosseous muscles were interconnected to the dorsal aponeurosis, extending to the fingertips and modeled as distinct muscle-tendon units. **g**, Foot, all four layers of intrinsic foot muscles individually modeled as muscle-tendon units. **h**, Detailed back musculature from Gray's Anatomy [1] (left) and the corresponding muscle-tendon units in the model (right).

Actor-Critic [50](SAC)-based controller, which included a hierarchical actor module, an action refinement module and a critic network. The hierarchical actor module contained two networks: the group action network generated actions at the muscle group level, and the unit action network provided state-dependent adjustment weights for individual muscle-tendon units. Following the hierarchical actor, the action refinement module further adjusted the output actions for movement control by applying adjustment weights to group actions. At the same time, the critic network evaluated the state-action pairs and updated the estimation of outcome Q-values. Rewards for reinforcement learning were designed based on whether the model behaviour matched the reference data or whether the task goal was achieved, as well as task-specific terms (see following sections).

The network controller was trained through a stage-wise learning strategy that progressively increased task complexity, analogous to the human motor learning process. This stage-by-stage learning approach facilitated exploration of the high-dimensional action space and achieved coordinated movement patterns within biomechanical constraints, providing a balance between computational efficiency and physiological validity.

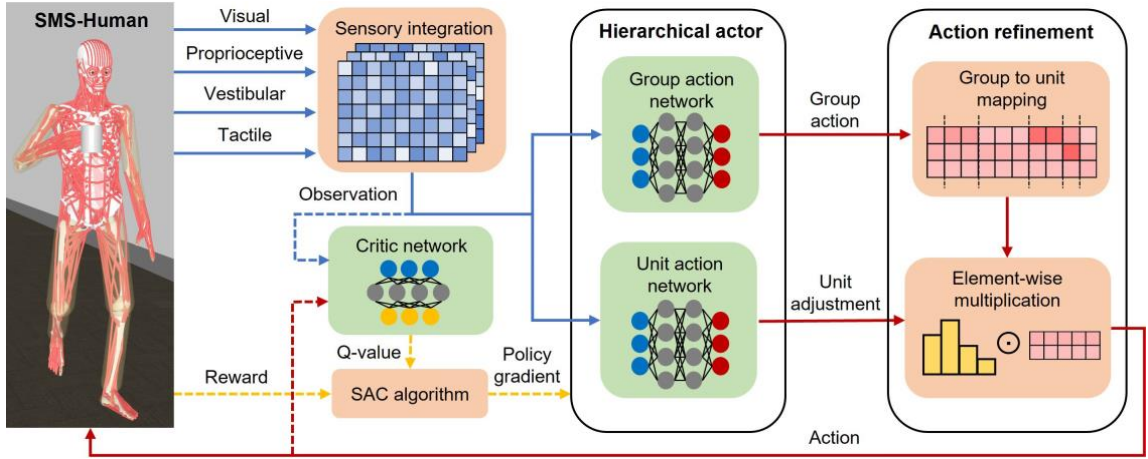

**Fig. 3 Schematic illustration of the hierarchical deep reinforcement learning for high-dimensional sensorimotor control.** Visual features and values from proprioceptive, vestibular and tactile sensors were integrated to form the observation matrix. The observation was then processed by a SAC-based controller including a critic network, and a hierarchical actor module followed by an action refinement module. The hierarchical actor module consisted of a high-level group action network generating shared actions for each muscle group, and a low-level unit action network producing state-dependent adjustment weights for each individual muscle-tendon unit. The action refinement module adjusted output actions at unit-level via group-unit mapping and element-wise multiplication of group actions and unit adjustment weights. The critic updated the estimation of Q-values, informing policy improvement.

## Simulation of bipedal walking

The DRL approach was used to train the network controller for simulating natural human bipedal walking by the SMS-Human model. During training, the network controller received kinematic references derived from motion capture data of body keypoint trajectories from an adult male performing bipedal walking, as well as the sensory signals generated by our SMS-Human model in the simulation environment. Three approaches were introduced during DRL of the controller to achieve the bipedal walking task. First, we used a stage-wise learning approach, in which the learning complexity was gradually increased during training, including increasingly stricter constraints on pose tracking and multi-joint coordination. Second, the DRL algorithm maximized a reward function that encouraged simulated movements to match the reference motion capture data, while maintaining muscle activations within the physiological range. Third, to promote control efficiency while preserving biomechanical validity, we implemented targeted simplifications by limiting the ROM of non-critical articulations of joints (e.g., finger joints) during bipedal walking.

The trained network controller successfully reproduced natural human bipedal walking patterns with a speed of 1.22 m/s and a gait cycle duration of 1.04 s (Fig. 4a and Supplementary Video 1). Quantitative evaluations indicated the high precision in the kinematics and dynamics of the simulated locomotion. First, joint angle trajectories of eight representative joints closely tracked the motion capture data from the human subject throughout three gait cycles of bipedal walking (Fig. 4b). Second, muscle activities in our SMS-Human model during simulated walking could largely match the EMG signals (with some exhibiting time-locked phase shifts) recorded from corresponding muscles of the same reference human subject during his bipedal walking (Fig. 4c). Furthermore, the spatiotemporal patterns of simulated muscle activities were in line with the expected functions of these muscles during walking. For example, gluteus maximus was activated during the early stance to drive hip extension, while rectus femoris was activated in loading response phase and mid-stance phase to stabilize the knee against ground reaction forces. Biceps femoris displayed activation during the swing phase, functioning to flex the knee joint. Vastus medialis was activated during the late swing for knee extension, preparing for heel strike. Tibialis anterior showed two activation peaks that coincided with the timing of EMG signals - in early stance and during swing, synchronizing with dorsiflexion demands. Peroneus longus exhibited subtle activation peak during the push-off, reflecting its role in assisting plantar flexion. Medial gastrocnemius and soleus coordinated propulsion, peaking in late stance during push-off. These results demonstrate that the learned network controller could generate muscle activations that closely match natural human activation patterns.

In Figure 4d, the comprehensive dynamics of joints and muscles is visualized to illustrate the capability of the SMS-Human model to simulate and analyze complex whole-body musculoskeletal coordination. Selective examination of specific components in our model could also help to capture subtle yet important aspects of walking. For example, we detected small angular movements (within ±0.02 radians) of the lumbosacral and cervicothoracic joints (Fig. 4e, upper panels), indicating appropriate torso stability. Notably, our simulation provided access to activation patterns of deep muscles that are difficult to measure experimentally, including the flexor digitorum longus in the posterior calf and the internal oblique in the abdomen (Fig. 4e, lower panels). The coordinated activation of these deep muscles aligned with their known anatomical functions - the deep toe flexors showing activity during foot push-off to aid propulsion and the internal oblique maintaining trunk stability throughout the gait cycle. These findings underscore the potential usefulness of our comprehensive modeling of the human musculoskeletal system for predicting detailed dynamics of all human muscles and joints in the absence of complete experimental measurements on humans.

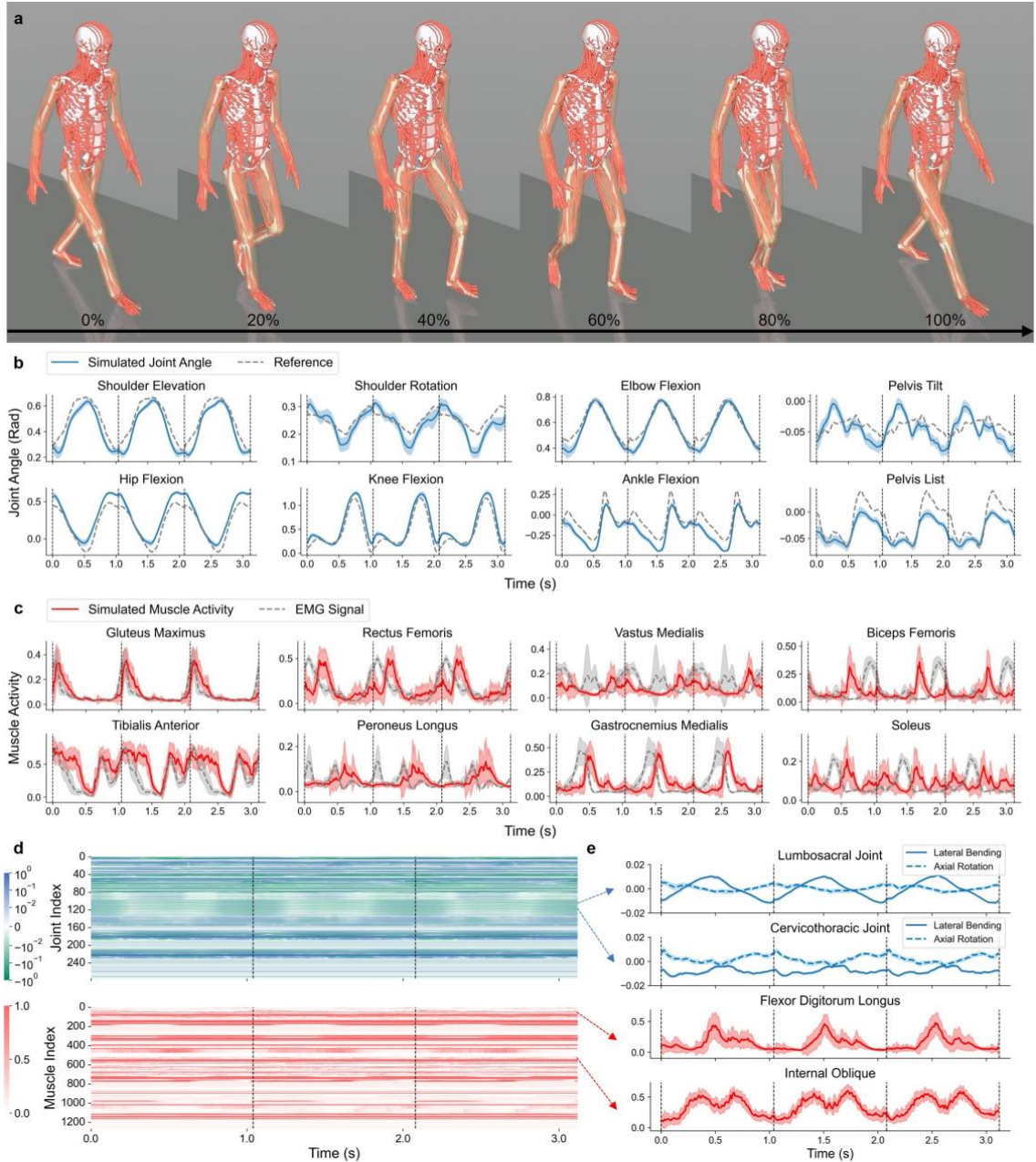

**Fig. 4 Whole-body simulation for bipedal locomotion**. **a**, Simulated bipedal locomotion sequence demonstrating a full gait cycle during straight walking. **b**, Comparison of SMS-Human-simulated (blue lines) and reference (gray dashed lines) joint angle trajectories for eight representative upper and lower body joints during walking, for three gait cycles. Pelvis tilt is the rotation around the lateral axis, and pelvis list is the rotation around the anteroposterior axis, both relative to the ground. **c**, Simulated muscle activity patterns (red lines) and experimental EMG signals (gray dashed lines) for eight major lower limb muscles during walking, for three gait cycles. **d**, Comprehensive joint kinematics and muscle activations during the three gait cycles of bipedal walking. Upper panel: temporal changes of joint angles (in radian) in an example walking trial; lower panel: simulated muscle activity profiles in the same walking trial. **e**, Selected examples of simulated joint angle and muscle activity during bipedal walking. Upper panel: lateral bending/axial rotation of two spinal joints showing subtle movements, indicating stable torso control during walking; lower panel: activity patterns of two deep muscles, which could not be readily measured experimentally. In **b-e**, vertical black dash lines denote the time of right heel strike; shaded regions denote one standard deviation (n=30 for simulated joint angles and muscle activities, n=7 for EMG).

## Vision-guided object manipulation

We next examined the ability of the SMS-Human model in simulating a visually guided manipulation task involving picking up an "object" (a bottle) on the table, moving it first to align with the "target" (a virtual bottle) above the table, and then following the target's random leftward or rightward movement (starting at 0.5 s after task initialization). Unlike bipedal walking, this task requires coordinated eye-head-hand movements, in which visual input plays a primary role in guiding appropriate motor execution. Multisensory integration further contributes key information for the states of the hand, arm, and object. Furthermore, this object manipulation task did not require human reference data. Instead, a specifically designed reward function was used to encourage the model to reduce both positional and orientational differences between the object and target during training, while achieving coordinated eye-head-hand movements.

Binocular images (224×224 pixels, 80° field-of-view) captured by cameras in eyes were pre-processed by a distance-dependent Gaussian filter to simulate foveal visual sensing with peripheral blurring. A pre-trained convolutional neural network was then applied to acquire task-relevant visual representations as inputs for the network controller. The coordinates and orientation of the object and target were not explicitly provided to the controller during task execution after training. The controller therefore must infer the spatial relationship and movement goal through its visual inputs and other sensory feedback.

The trained network controller successfully performed the bottle pick-up and translocation task in an eye-head-hand coordinated manner (Fig. 5a and Supplementary Video 2). Notably, the model exhibited visual tracking behaviour with coordinated eye-head movements to fixate around the object (Fig. 5a). The learned visuomotor control of our model was similar to that of a human subject in performing the same task. The eyes of the subject were elevated to guide the hand movement during bottle pick-up and rotated laterally with the translocating object (Fig. 5b).

Quantitative evaluation indicates that SMS-Human model achieved expected object manipulation towards the target within 0.5 s before target moving. During the target tracking phase (after 0.5 s), the object stayed close to the target with an average translational difference of 0.15±0.25 cm and a rotational difference of 0.01±0.02 radians, while the average fixation-to-object offset was 0.53±0.43 cm (mean±SEM, n=10, Fig. 5c). The convergence of position and orientation of the object to the target (Fig. 5d) suggests effective vision-guided motor control, in which the model continuously updated its motor commands based on visual and other sensory feedback.

The movement trajectories in Fig. 5e illustrate human-like task-dependent eye-head-hand kinematic patterns. During bottle pick-up, both eyes exhibited synchronized elevation movements, with upward rotation angles increasing rapidly and stabilizing thereafter. The head elevation angle showed a corresponding temporal profile, establishing a stable foundation for visual tracking. During the target tracking phase, the model demonstrated direction-specific eye movement patterns. For rightward target movement, the left eye exhibited gradual adduction toward the medial while the right eye showed abduction toward the right lateral. This pattern was mirrored during leftward target movement, demonstrating correct eye orientation for object tracking. Concurrently, the model executed proper arm movements, characterized by coordinated adjustments in humerus elevation and rotation, and radius pronation to maximize spatial alignment between the object and the target. Throughout the process, the model maintained stable eye and head elevation while executing smooth trajectory control of upper-limb joints, demonstrating robust manipulation capability during moving target pursuit.

The learned activation patterns of extraocular muscles during object manipulation are shown in Fig. 5f, representing the unique capability of our model to simulate natural control of eye movements. Bilateral superior rectus muscles exhibited strong initial activation to drive eyeball elevation, followed by sustained moderate activity, working in concert with inferior rectus muscles to maintain vertical eye position. The superior and inferior oblique extraocular muscles exhibited complementary activation patterns that fine-tuned the eye orientation. During the dynamic tracking phase, distinct direction-dependent extraocular muscle activation patterns were triggered by model's visual feedback of the target movement. For rightward tracking, the left medial rectus and right lateral rectus showed enhanced activation, while the activities of their respective antagonists (left lateral rectus and right medial rectus) were reduced, resulting in the shift of the fixation point to the right. Reversed patterns were observed for leftward tracking. Thus, our model was capable of generating appropriate synergistic muscle activation for object tracking, and this visual tracking was integrated with other sensory signals to coordinate motor behaviours.

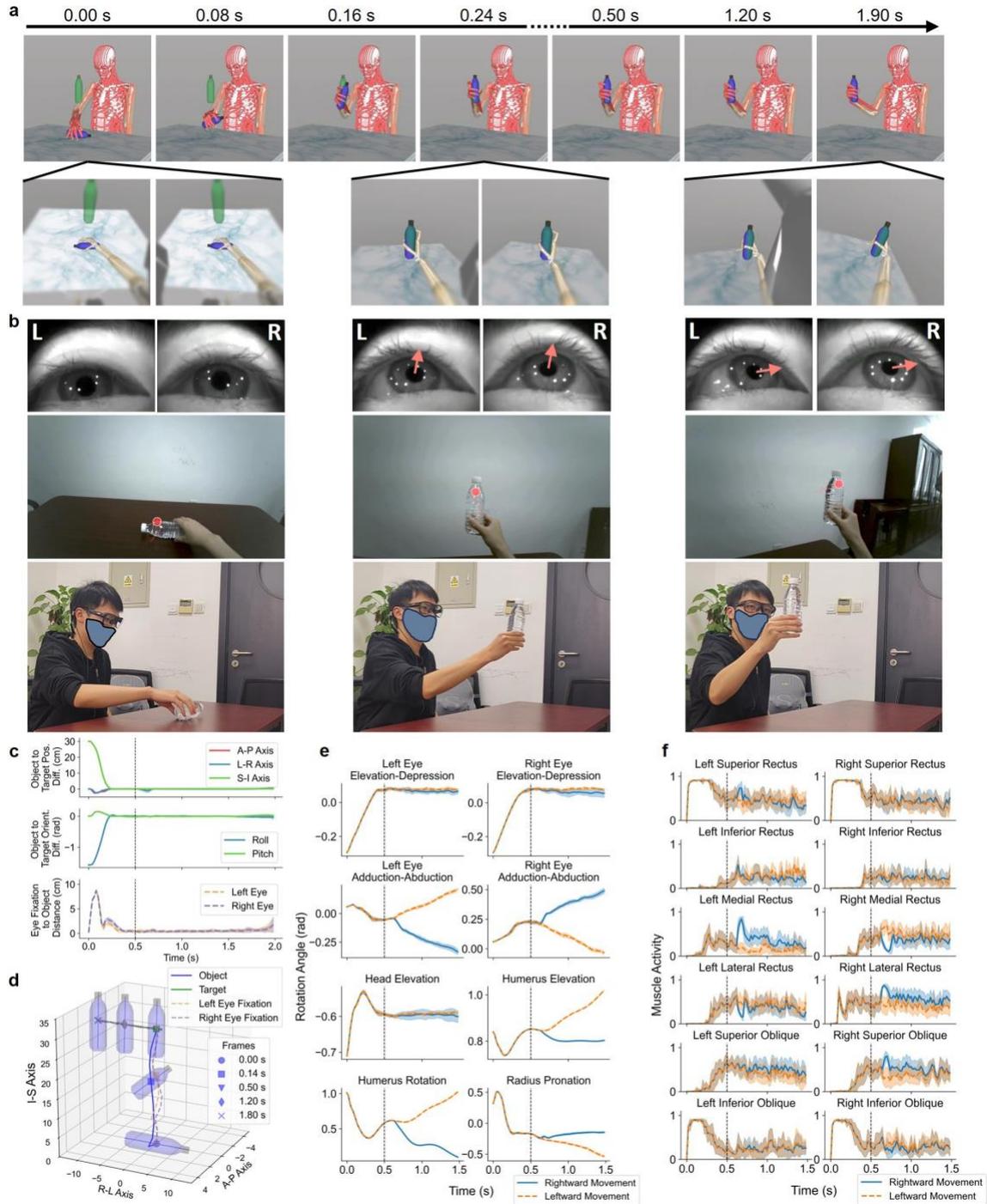

**Fig. 5 Visuomotor coordination in the object manipulation task. a**, Top: Simulated model behaviour showing picking-up, holding, and translocating the object (blue bottle) in accordance to the target cue (semitransparent green bottle). The object and the target cue largely overlapped after 0.24 s. Bottom: Visual inputs to the model at three time points during the task. **b**, The same task performed by a representative human subject, showing the eye-head-hand coordination at three key time points during the task. Top panels: human eye movement directions. Red arrows, the combined eye rotation directions. Middle panels: human visual scenes and fixation points. Bottom panels: human behaviour recorded from a camera in the environment. **c**, Spatial differences between the object and target, and between the eye fixation point and object. Zero value indicates exact alignment. Top panel: position differences between centers of the object and target. Orthogonal coordinate axes of the model: A-P, anterior-posterior; L-R, left-right; S-I, superior-inferior. Middle panel: orientation differences between the object and target, measured in two angular axes. Bottom panel: distance between the eye fixation points and the object. **d**, 3D trajectories (in cm) during target tracking of the rightward-moving task, with the object's initial position at the origin of coordinates. **e,f**, Angular kinematics of eye rotations, head and upper limb articulations (**e**), and activation of bilateral extraocular muscles (**f**) in our model during object manipulation task, for leftward (orange) and rightward (blue) movements. In **c-f**, shaded regions denote one standard deviation across 10 trials; vertical black dashed lines denote target movement onset time; A-P, anterior-posterior; L-R, left-right; S-I, superior-inferior.

## Bicycle riding

To demonstrate the capability of our SMS-Human model for simulating human-machine interaction, we designed a bicycling task that requires coordinated whole-body musculoskeletal control for interaction with a bicycle, which was confined in vertical orientation with movable pedals and wheels for forward movement. During training, the network controller was rewarded for maintaining stable upper body posture while holding the fixed handlebar via sensory feedback signals and for matching the model's body keypoint trajectories to synthetic 3D trajectories from pre-generated bicycling motion (Fig. 6a).

During the task performance after training, the model successfully executed rhythmic pedaling movements while maintaining upper body stability, achieving a natural cycling posture (Supplementary Video 3). The sequential visualization of the pedaling cycle showed smooth transitions between the top and bottom foot positions characteristic of proper cycling mechanics (Fig. 6b). Quantitative analysis reveals high precision in the control of simulated body movements, with a maximum spatial tracking error of 2.38±0.27 cm (mean±SEM, n=10) for body keypoints across all three axes (Fig. 6c). The temporal profiles of lower limb kinematics demonstrate coordinated cyclical patterns similar to human cycling motion capture experiments[51]. The vertical contact force as measured by the tactile sensors at the feet showed characteristic peaks during the power phase of each cycle, indicating effective force transmission to the bicycle (Fig. 6d). These results indicate that our model is capable of simulating coordinated whole-body movements while interacting with external devices.

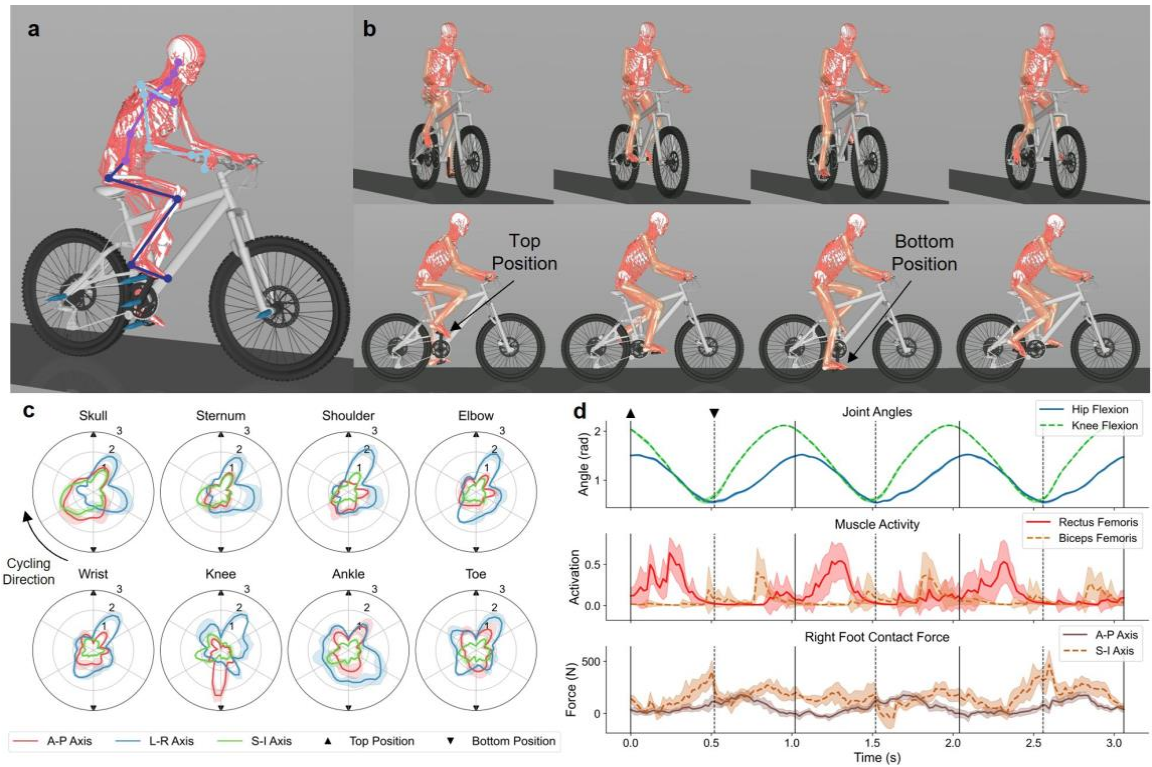

**Fig. 6 Whole-body simulation for bicycling. a**, Simulation of a bicycling task by the SMS-Human model, with body keypoints of the right side illustrated in different colors for different body parts (body keypoints of the left side are not shown for visualization clarity). The network controller was rewarded for matching the motion trajectories of these keypoints to the synthetic reference trajectories during training. The blue arrows represent the rotation axes of the crankset, pedals, and wheels of the bicycle. **b**, Sequential visualization of the pedaling cycle of learned model behaviour. Top and bottom positions are marked for the right foot. **c**, Spatial deviations (in cm) of representative body keypoints to their reference positions (with 0 value indicating exact matching), in each pedaling cycle during task execution after training, along three axes of the environment. **d**, Temporal patterns of lower limb kinematics and dynamics during cycling, including representative joint angles (top panel), activation of muscles (middle panel) and contact force between right foot and the pedal (bottom panel). Black solid and dashed lines indicate the top and bottom positions of the right foot respectively. In **c-d**, shaded regions denote one standard deviation across 10 trials; A-P, anterior-posterior; L-R, left-right; S-I, superior-inferior.

# Discussion

Coordinated human movements arise from the intricate integration of multimodal sensation, sensorimotor transformation, and motor execution within a closed-loop system. This work seeks to computationally realize these processes and generate adaptive motor behaviours across various tasks with a comprehensive whole-body sensory-musculoskeletal model. By coupling multisensory inputs with high-dimensional motor outputs through learning-based approaches, our model provides a new platform for investigating the spatiotemporal dynamics of the human musculoskeletal system during motion.

A central focus of this work was achieving the most detailed and precise musculoskeletal representation to date, comprising 206 bones, 278 joints, and 1,266 muscle-tendon units. Each component was thoroughly parameterized using previously published anatomical and physiological data, allowing for a structurally accurate simulation of human body dynamics. Beyond anatomical completeness, the model integrates sensory modalities, including visual, vestibular, proprioceptive, and tactile inputs, enabling closed-loop sensorimotor control. This integration allows the model to generate realistic movement patterns by continuously adapting motor outputs based on multimodal sensory feedback.

Our hierarchical deep reinforcement learning method effectively addressed the inherent challenges of controlling high-dimensional motor systems. The stage-wise learning strategy, inspired by human motor learning process, allowed our model to progressively tackle tasks of increasing difficulty. The integration of stage-wise learning with hierarchical action space representation makes the high-dimensional control problem more tractable, by elevating sampling efficiency and accelerating convergence of the learning process. The capability of performing coordinated movements with high physiological and behavioural fidelity represents a significant technical advance in physics-based human movement simulation. The SMS-Human model could simulate muscle activation dynamics for studying human motor learning and control, which is supported by the strong agreement between our simulation results and experimental data on natural human movements. We could also provide fine-grained dynamic profiles for all motion-related muscles, many of which are not feasible to be measured directly in behaving humans.

The SMS-Human model holds broad implications across various fields. In rehabilitation engineering, it could facilitate studies of the sensory feedback-dependent motor adaptation, the effects of sensory deficits on motor learning, and the structural and dynamic limits of the musculoskeletal system related to injury. The model can thus support the design and optimization of rehabilitation protocols, prosthetics, and therapeutic interventions by simulating patient-specific responses and compensatory strategies. In sports science, our model could offer a platform for optimizing movement and reducing injury risks by analyzing internal biomechanical loads and muscle activations. For human–robot interaction, the model could inform the development of control strategies aligned with human sensorimotor capabilities. Furthermore, the integration of sensory and musculoskeletal systems lays a foundation for developing humanoid robots with robust structural design, multimodal sensing, and improved motor performance.

Despite the comprehensive anatomical details in our model, several limitations exist. First, the Hill-type muscle model we employed does not capture the full biological complexity of muscle behaviours, resulting in simplified muscle-tendon dynamics. Second, the sensory systems in the model remain a simplified version of biological systems. Third, the available human EMG data are limited, thus preventing a more thorough comparison of the dynamics of muscle activation between our model's prediction and actual human data. Future work could further improve the fidelity in sensory-musculoskeletal representation and broaden the range of movement tasks in more complex environments.

In this study, we present a high-dimensional control framework for simulating human motor behaviours and establish a useful benchmark for addressing the challenges of high-dimensional learning and control. This framework provides a novel computational platform for exploring the embodiment of human intelligence and enables large-scale simulations of motor control with multisensory feedback. Future research may explore more advanced learning and control algorithms, incorporate higher-order neural processes involving reasoning and decision-making, and extend the model's applicability across diverse domains.